\documentclass[twocolumn,showpacs,preprintnumbers,draftclsnofoot]{revtex4}
\usepackage{amsmath}
\usepackage{amssymb}
\usepackage{graphicx}
\usepackage{dcolumn}
\usepackage{bm}
\usepackage{amssymb}
\usepackage{graphicx}
\usepackage{dcolumn}
\usepackage{bm}

\begin{document}

\title{ Localized Floquet states in gated bilayer graphene induced by a focused optical beam with orbital angular momentum  }
\author{Ma Luo\footnote{Corresponding author:swym231@163.com} }
\affiliation{School of Optoelectronic Engineering, Guangdong Polytechnic Normal University, Guangzhou 510665, China}

\begin{abstract}

We theoretically studied the Floquet state of gated bilayer graphene, which is irradiated by normally incident focused Gaussian beam with orbital angular momentum (OAM). According to the Floquet theory, in-plane and out-of-plane electric field of the OAM beam periodically perturbs the intralayer and interlayer hopping, which are equivalent to the presence of effective staggered sublattice potential and next-nearest neighboring interlayer hopping within the light spot region, respectively. The combination of the effective terms and the gated voltage form an effective trapping potential, which hosts localized quantum states with energy level being within the energy gap of the non-irradiated gated bilayer graphene. The energy spectrum can be tuned by the amplitude of the optical beam and additional static magnetic field. By engineering the parameters, valley-polarized two-fold degenerated zero energy Floquet states can be induced.

\end{abstract}

\pacs{00.00.00, 00.00.00, 00.00.00, 00.00.00}
\maketitle

\section{Introduction}

Periodic perturbation on a quantum system could drive the quantum states into photonic dressed states, which is described by the Floquet theory \cite{Rodriguez08,Takashi09,Savelev11,ArijitKundu14,Taboada17,Taboada171,DalLago15,MichaelVogl19,MichaelVogl20}. In the presence of optical irradiation on two-dimensional material, such as graphene, the electrons are periodically perturbed, so that the band structure of the Bloch states are modified. The Floquet band structures can be experimentally measured by the method of angle-resolved photoemission spectroscopy (ARPES) \cite{YHWang13,Farrell16,YaoWang18,HaifengYang18,BaiqingLv19}. If the incident optical field has circular polarization, varying types of topological phases with nontrivial bulk band gap can be induced, such as Floquet topological insulator (TI) and Chern insulator \cite{Inoue10,Kibis10,Calvo11,ShuTing16,Mukherjee18,Yunhua17,Ledwith18,HangLiu18,LongwenZhou18,maluo21F}. In bilayer graphene (BLG), combination of gate voltage and irradiation can induce valley-polarized metallic phase and spin-valley polarized quantum anomalous Hall phase \cite{XuechaoZ20,XikuiMa21,maluo21}. One of the most exciting feature of the topological phases is the presence of topological edge states that are spatially localized at open boundaries \cite{Piskunow14,Usaj14,Claassen16,Tahir16,Puviani17,Hockendorf18,PerezPiskunow15,AaronFarrell15,AaronFarrell16}. The topological edge states also appear at interface between two regions of a piece of graphene under different type of irradiating condition \cite{Arijit16,HuamanGonzalo19}. The topological edge states could be applied to construct on-chip photoconductive devices and opto-spintronic devices \cite{JWMcIver20}. On the other hand,  combination of inter-layer twisting and irradiation can induced flat topological Floquet-Bloch bands, which could host strongly correlated nonequilibrium quantum states \cite{twist1,twist2,twist3,twist4,twist5}.

\begin{figure}[tbp]
\scalebox{0.58}{\includegraphics{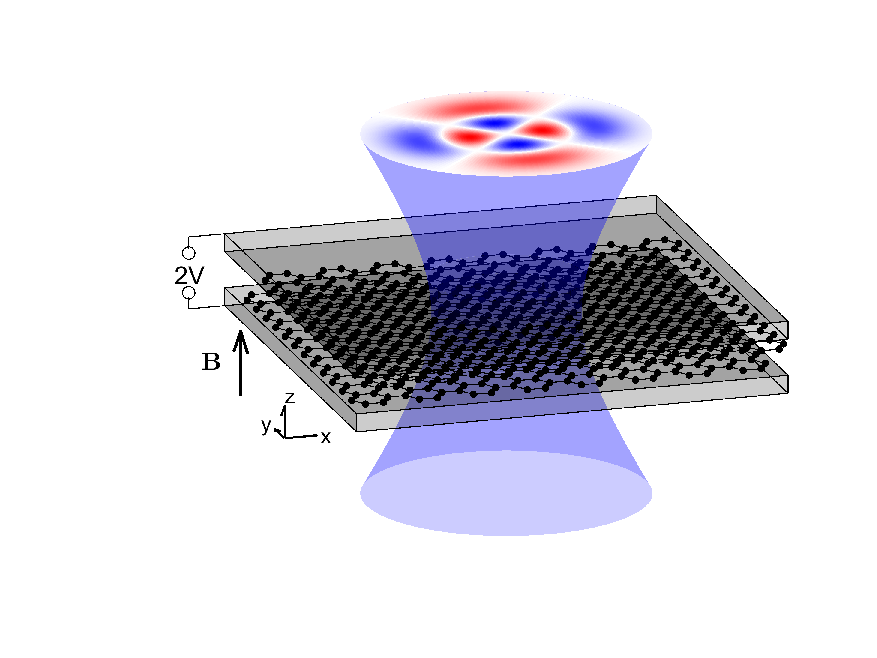}}
\caption{ The scheme of bulk BLG irradiated by tightly focused OAM beam. The BLG is clamped by top and bottom transparent electrodes, which induce voltage difference of $2V$ between the top and bottom graphene layers. An external static magnetic field along the out-of-plane direction $\mathbf{B}$ is applied. The spatial dependent of the amplitude of the in-plane electric field of the OAM beam with $l=2$ and $p=2$ is plotted at the top cross section of the beam.   }
\label{figure_0}
\end{figure}

Because the wavelength of the optical field is much larger than the de Broglie wavelength of the electrons, the optical field is usually approximated as uniform oscillating electric field. However, recent development of optical metalens enable tightly focusing of optical field in sub-wavelength scale \cite{metalen11,metalen12,metalen16,metalen18a,metalen18b}. In this paper, bulk bilayer graphene that is irradiated by tightly focused Gaussian beam, as shown by the schematic plot in Fig. \ref{figure_0}, are studied. For the electron with energy level being near to the Fermi level of pristine graphene, the quantum states can be described by the continuous Dirac fermion model. In the presence of the optical irradiation, the corresponding Floquet state can be described by the continuous Floquet-Dirac fermion model \cite{Kitagawa11,Goldman14,Grushin14}. If the energy of the photon satisfies non-resonant condition, the high-frequency expansion of the continuous Floquet-Dirac fermion model is applied to obtained the effective Hamiltonian. For Gaussian beam with orbital angular moment (OAM), the vectorial optical field contains both in-plane and out-of-plane components electric field \cite{AllenLembessis96}. For bilayer graphenes (BLGs), the intra-layer and inter-layer hopping are periodically perturbed by the in-plane and out-of-plane components electric field, respectively. If the in-plane component electric field has circular polarization, a valley-dependent effective staggered sublattice potential is induced in each graphene layer. The combination of the in-plane and out-of-plane perturbation induces effective next-nearest-neighboring (NNN) inter-layer hopping terms. Because the amplitude and phase of the optical field is spatially dependent, the effective staggered sublattice potential and the effective NNN inter-layer hopping terms are also spatially dependent.

In this paper, we engineered the combination of the OAM beam irradiation, gate voltage and static magnetic field on a BLG, which can induce a valley-polarized localized Floquet states with zero energy. For bilayer graphene (BLG), the gate voltage along the perpendicular direction induces a bulk band gap. In the presence of the OAM beam, the spatial dependent effective staggered sublattice potential has large magnitude within the region of the light spot, which can change the size and sign of the bulk band gap. Because the stagger sublattice potentials of the two valleys have opposite sign, the irradiation effectively induces a valley-polarized trapping potential within the region of the light spot. The energy spectrum of the localized Floquet states is dependent on the amplitude as well as the radial and azimuthal indices of the OAM beams. Because of the inversion symmetry of the irradiation effect, each energy level is four-fold degenerated. As the amplitude of the OAM beam reaches some critical values, two four-fold degenerated energy levels in one valley reach zero, while another valley remain being gapped, so that eight-fold degenerated valley-polarized zero energy Floquet states are induced. The degeneration can be broken by additional out-of-plane static magnetic field. By tuning the static magnetic field, valley-polarized two-fold degenerated zero energy Floquet states can be induced.

The article is organized as follows: In Sec. II, the high-frequency expansion of the continuous Floquet-Dirac fermion model is applied to deduce the effective Hamiltonian and Schr$\ddot{o}$dinger equation in real space. In Sec. III, the numerical result of the energy spectrum is given. The scheme to obtain zero energy Floquet states is discussed. In Sec. IV, the conclusions are given.

\section{Theoretical model}

In gated BLG, the quantum state of the electron near to the Fermi level can be described by the continuous Dirac fermion model. The model Hamiltonian is given as
\begin{equation}
H=\left[\begin{array}{cccc}
V & \hbar v_{F}k^{\tau}_{-} & 0 & 0 \\
\hbar v_{F}k^{\tau}_{+} & V & t_{\bot} & 0 \\
0 & t_{\bot} & -V & \hbar v_{F}k^{\tau}_{-} \\
0 & 0 & \hbar v_{F}k^{\tau}_{+} & -V \\
\end{array}\right]\label{HamiltonianD1}
\end{equation}
where $k^{\tau}_{\pm}=\tau k_{x}\pm ik_{y}$ with $\tau=\pm1$ being the valley index of the K and K$^{\prime}$ valleys, $k_{x}$ and $k_{y}$ are the wave number relative to the K or K$^{\prime}$ point, $2V$ is the potential difference between the top and bottom layers due to the gated voltage, $t_{\bot}=-0.39$ eV is the hopping constant between nearest-neighboring inter-layer lattice sites, and $v_{F}$ is the Fermi velocity of graphene. The value of $\hbar v_{F}$ is given as $\hbar v_{F}=3t_{\parallel}a/2$ with $t_{\parallel}=-2.8$ eV and $a=0.142$ nm being the hopping constant and the distance between the nearest-neighboring intra-layer lattice sites, respectively. In the presence of optical field, which is described by a periodically oscillating vector potential $\mathbf{A}(\mathbf{r},t)$, the electrons are periodically perturbed, so that the Hamiltonian becomes time dependent. Assuming that the spatial derivative of $\mathbf{A}(\mathbf{r},t)$ has small magnitude, the wave functions of the electrons along the in-plane direction can be approximated as Bloch states with spatially varying amplitude. By applying the dipolar approximation for the intra-layer hopping of the time-dependent Hamiltonian, the wave numbers $k_{x}$ and $k_{y}$ are replaced by the Peierls substitution as $k_{x}+e_{0}A_{x}(\mathbf{r},t)/\hbar$ and $k_{y}+e_{0}A_{y}(\mathbf{r},t)/\hbar$, respectively. Along the out-of-plane direction, the inter-layer hopping is periodically perturbed by the out-of-plane component of the optical field, which is described by the time-dependent Peierls phase in $t_{\bot}$ \cite{Peierls33}. Thus, the hopping terms are replaced as $t_{\bot}e^{\pm\frac{i2\pi d_{z}A_{z}(\mathbf{r},t)}{\Phi_{0}}}$, with $d_{z}=0.34$ nm being the distance between the two graphene layers, $\Phi_{0}=\pi\hbar/e_{0}$ being the magnetic flux quantum, and the $\pm$ sign representing upward or downward hopping.

For OAM beam, the vector potential is spatially dependent. Because $d_{z}$ is much smaller than the wavelength of the OAM beam, the optical fields at the top and bottom layer are nearly the same. The BLG is suspended at the x-y plane with $z=0$. The OAM beam propagates along z axis with minimum beam waist at $z=0$. Thus, under the irradiation of OAM beam with radial and azimuthal indices being $p$ and $l$, the vector potential at the BLG is given as
\begin{equation}
\mathbf{A}(\mathbf{r},t)=[(\hat{x}+i\eta\hat{y})U-\frac{ie^{i\eta\phi}}{k}(\frac{\partial U}{\partial r}+\frac{\eta l}{r}U)\hat{z}]e^{-i\omega t}+c.c. \label{fieldOAM}
\end{equation}
, where $U(\mathbf{r})=u(r)e^{-il\phi}$ and
\begin{equation}
u(r)=A_{0}(\frac{\sqrt{2}r}{w_{0}})^{l}L_{p}^{l}(\frac{2r^{2}}{w_{0}^{2}})e^{-\frac{r^{2}}{w_{0}^{2}}}
\end{equation}
, with $A_{0}$ being the amplitude of the optical field, $\eta=\pm$ representing left and right circular polarization, $r$ and $\phi$ being the cylindrical coordinate of $\mathbf{r}$ at the x-y plane, $\omega$ being frequency of the optical field, $k=\omega/c$ being the wave number, $w_{0}$ being the beam waist at $z=0$, and $L_{p}^{l}(\xi)$ being the associated Laguerre's Polynomial of $\xi$. Inserting $\mathbf{A}$ into the Hamiltonian gives the time-dependent Hamiltonian $H(t)$. The time-dependent matrix elements, $H_{1,2}(t)$ and $H_{3,4}(t)$, are given as
\begin{equation}
H_{1,2(3,4)}(t)=\frac{e_{0}A_{0}}{\hbar}u(r)[(\tau+\eta)e^{-il\phi-i\omega t}+(\tau-\eta)e^{il\phi+i\omega t}]
\end{equation}
; $H_{2,1}(t)$ and $H_{4,3}(t)$, are given as
\begin{equation}
H_{2,1(4,3)}(t)=\frac{e_{0}A_{0}}{\hbar}u(r)[(\tau-\eta)e^{-il\phi-i\omega t}+(\tau+\eta)e^{il\phi+i\omega t}]
\end{equation}
The time-dependent interlayer hopping term $H_{2,3(3,2)}(t)$ can be expanded as
\begin{eqnarray}
&&H_{2,3(3,2)}(t)=t_{\bot}e^{\zeta\frac{i2\pi d_{z}A_{z}(\mathbf{r},t)}{\Phi_{0}}}=t_{\bot} \\
&&\sum_{m=-\infty}^{+\infty}{J_{m}\{-\zeta\frac{4\pi d_{z}A_{0}}{k\Phi_{0}}[\frac{\partial u(r)}{\partial r}+\frac{\eta l}{r}u(r)]\}e^{im(l-\eta)\phi+im\omega t}}\nonumber
\end{eqnarray}
, where $\zeta=1$ and $\zeta=-1$ are for $H_{2,3}(t)$ and $H_{3,2}(t)$, respectively; $J_{m}(\xi)$ is the m-th order Bessel function of variable $\xi$ \cite{Calvo13}. As a result, the time-dependent Hamiltonian can be expanded as $H(t)=\sum_{m=-\infty}^{+\infty}H_{m}e^{im\omega t}$. Considering the high frequency situations with $\hbar\omega>2|t_{\parallel}|$, the effective Floquet Hamiltonian can be given as \cite{Kitagawa11,Goldman14,Grushin14}
\begin{equation}
H^{eff}=H_{0}+\sum_{m>0}{\frac{[H_{+m},H_{-m}]}{m\hbar\omega}}+\mathcal{O}(\frac{1}{\omega^{2}})
\end{equation}
Performing the high-frequency expansion, the effective Floquet Hamiltonian is given as
\begin{equation}
H^{eff}=\left[\begin{array}{cccc}
V+\Delta & \hbar v_{F}k^{\tau}_{-} & t_{\bot,1} & 0 \\
\hbar v_{F}k^{\tau}_{+} & V-\Delta & t_{\bot,0} & -t_{\bot,1} \\
t_{\bot,1}^{*} & t_{\bot,0} & -V+\Delta & \hbar v_{F}k^{\tau}_{-} \\
0 & -t_{\bot,1}^{*} & \hbar v_{F}k^{\tau}_{+} & -V-\Delta \\
\end{array}\right]\label{HamiltonianD1e}
\end{equation}
, where $\Delta=-4\tau\eta[e_{0}A_{0}v_{F}u(r)]^{2}/(\hbar\omega)$, $t_{\bot,0}=t_{\bot}J_{0}\{\frac{4\pi d_{z}A_{0}}{k\Phi_{0}}[\frac{\partial u(r)}{\partial r}+\frac{\eta l}{r}u(r)]\}$, $t_{\bot,1}=[e_{0}A_{0}v_{F}t_{\bot}/(\hbar\omega)]u(r)J_{1}\{\frac{4\pi d_{z}A_{0}}{k\Phi_{0}}[\frac{\partial u(r)}{\partial r}+\frac{\eta l}{r}u(r)]\}[(\tau-\eta)e^{i\eta\phi}+(\tau+\eta)e^{-i\eta\phi}]$. The terms $\Delta$, $t_{\bot,0}$ and $t_{\bot,1}$ are dependent on the spatial location $(x,y)$, or the equivalent polar coordinates $(r,\phi)$. For non-irradiated BLG, two energy bands have parabolic dispersion around zero energy, and the absolute value of the other two energy bands are above the energy level $|t_{\bot}|$. In order to study the low-energy states, whose energy levels are much smaller than $|t_{\bot}|$, the four-band model in Eq. \ref{HamiltonianD1} can be projected into a two-band model \cite{EdwardMcCann06}. For the irradiated BLG, $|t_{\bot,0}|$ is smaller than $|t_{\bot}|$, because $|J_{0}(\xi)|<1$. At the radial location $r_{m}$ that corresponds to the maximum value of $|\frac{\partial u(r)}{\partial r}+\frac{\eta l}{r}u(r)|$ within the spatial region, $|t_{\bot,0}|$ has minimum value. The first zero point of $J_{0}(\xi)$ is $\xi_{0}\approx2.4048$, i.e., $J_{0}(\xi_{0})=0$. As $A_{0}$ reaches the critical value $A_{0,\bot}$ that satisfies the condition $\frac{4\pi d_{z}A_{0,\bot}}{k\Phi_{0}}|\frac{\partial u(r)}{\partial r}|_{r=r_{m}}+\frac{\eta l}{r_{m}}u(r_{m})|=\xi_{0}$, $t_{\bot,0}$ is equal to zero. In this case, the two-band model is invalid, because all four bands have low energy. For the case with $A_{0}\ll A_{0,\bot}$, $|t_{\bot,0}|$ is slightly smaller than $|t_{\bot}|$, so that the effective Hamiltonian in Eq. \ref{HamiltonianD1e} can be projected into the two-band model. The value of $A_{0,\bot}$ is dependent on only four parameters of the OAM beam, which are $\omega$, $w_{0}$, $l$ and $p$. Our numerical results found that the Floquet states with zero energy level appear in the parameter regime with $A_{0}\approx A_{0,\bot}$ or $A_{0}>A_{0,\bot}$, so that we do not apply the two-band model.

In the additional presence of a static uniform magnetic field $\mathbf{B}^{s}=B_{x}\hat{x}+B_{y}\hat{y}+B_{z}\hat{z}$, the static vector potential $\mathbf{A}^{s}=A^{s}_{x}\hat{x}+A^{s}_{y}\hat{y}=(zB_{y}-yB_{z}/2)\hat{x}+(-zB_{x}+xB_{z}/2)\hat{y}$ is inserted into the Peierls substitution, so that the matrix elements $H^{eff}_{1,2(3,4)}$ and $H^{eff}_{2,1(4,3)}$ contain addition terms $\tau e_{0}v_{F}A^{s}_{x}+ie_{0}v_{F}A^{s}_{y}$ and $\tau e_{0}v_{F}A^{s}_{x}-ie_{0}v_{F}A^{s}_{y}$, respectively. Because the inter-layer distance between the two graphene layers is small, the in-plane magnetic field has negligible effect on the energy spectrum. Thus, we only studied the effect of out-of-plane magnetic field, i.e.  $\mathbf{B}^{s}$ with $B_{x}=B_{y}=0$.

Because the matrix elements of the effective potential are spatially dependent, the translation symmetric of the Hamiltonian is broken. Thus, the momentum operators need to be expressed in real space as $\mathbf{k}=-i\nabla$, i.e. $k^{\tau}_{\pm}=-i\tau\frac{\partial}{\partial x}\pm\frac{\partial}{\partial y}$ is inserted into the effective Hamiltonian. The wave function of the electrons is a double pseudo-spinor field with four components, i.e., $\Psi=[\psi_{T,A},\psi_{T,B},\psi_{B,A},\psi_{B,B}]^{T}$. For each component, the first index indicates the top and bottom layer, and the second index indicates the A and B sublattice. The Schr$\ddot{o}$dinger equation of the effective Hamiltonian become eigenvalue problem of a system of partial differential equations of the four components, which is given as
\begin{widetext}
\begin{eqnarray}
(V+\Delta)\psi_{T,A}+[\hbar v_{F}(-i\tau\frac{\partial}{\partial x}-\frac{\partial}{\partial y})+\tau e_{0}v_{F}A^{s}_{x}+ie_{0}v_{F}A^{s}_{y}]\psi_{T,B}+t_{\bot,1}\psi_{B,A}=E\psi_{T,A}\nonumber \\
(\hbar v_{F}(-i\tau\frac{\partial}{\partial x}+\frac{\partial}{\partial y})+\tau e_{0}v_{F}A^{s}_{x}-ie_{0}v_{F}A^{s}_{y})\psi_{T,A}+(V-\Delta)\psi_{T,B}+t_{\bot,0}\psi_{B,A}-t_{\bot,1}\psi_{B,B}=E\psi_{T,B}\nonumber \\
t_{\bot,1}^{*}\psi_{T,A}+t_{\bot,0}\psi_{T,B}+(-V+\Delta)\psi_{B,A}+[\hbar v_{F}(-i\tau\frac{\partial}{\partial x}-\frac{\partial}{\partial y})+\tau e_{0}v_{F}A^{s}_{x}+ie_{0}v_{F}A^{s}_{y}]\psi_{B,B}=E\psi_{B,A}\nonumber \\
-t_{\bot,1}^{*}\psi_{T,B}+[\hbar v_{F}(-i\tau\frac{\partial}{\partial x}+\frac{\partial}{\partial y})+\tau e_{0}v_{F}A^{s}_{x}-ie_{0}v_{F}A^{s}_{y})\psi_{B,A}+(-V-\Delta)\psi_{B,B}=E\psi_{B,B} \label{equationgroup}
\end{eqnarray}
\end{widetext}
The system of partial differential equations in Eq. (\ref{equationgroup}) is numerically solved by applying finite different method. Because the in-gap states are localized within the beam radius, the computational domain is truncated at open boundaries, which is far away from the center of the OAM beam. In the absence of the static magnetic field, the Hamiltonian has inversion symmetry in both layers, so that the energy levels are four-fold degenerated. The presence of the static magnetic field breaks the inversion symmetry, so that the four-fold degenerated energy level split into four energy levels. Because the particle-hole symmetry is preserved, the energy levels above and below zero energy are symmetric.

Due to rotational symmetry, the wave functions of the eigen states have fixed angular momentum index $q$. In cylindrical coordinate, the spatial differential terms in Eq. (\ref{equationgroup}) are replaced as $-i\tau\frac{\partial}{\partial x}\pm\frac{\partial}{\partial y}=-i\tau e^{\pm i\tau\phi}\frac{\partial}{\partial r}\pm\frac{e^{\pm i\tau\phi}}{r}\frac{\partial}{\partial\phi}$. Because of the phase factor $e^{\pm i\tau\phi}$ in the matrix element $H^{eff}_{1,2(3,4)}$ and $H^{eff}_{2,1(4,3)}$, the angular momentum indices of the wave functions of the two sublattices in each graphene layer are different by the value of $\tau$. The matrix elements $H^{eff}_{2,3(3,4)}$ and $H^{eff}_{3,1(4,2)}$ have a phase factor $[(\tau-\eta)e^{i\eta\phi}+(\tau+\eta)e^{-i\eta\phi}]$. In the case that $\tau=\eta$, the phase factor is $2\tau e^{-i\eta\phi}$; in the case that $\tau=-\eta$, the phase factor is $2\tau e^{i\eta\phi}$. Thus, the phase factor can be written as $2\tau e^{-i\tau\phi}$. As a result, the four components of the wave functions $\Psi$ can be expressed as $\psi_{T,A}=\varphi_{T,A}(r)e^{i(q-\tau)\phi}$, $\psi_{T,B}=\varphi_{T,B}(r)e^{iq\phi}$, $\psi_{B,A}=\varphi_{B,A}(r)e^{iq\phi}$, and $\psi_{B,B}=\varphi_{B,B}(r)e^{i(q+\tau)\phi}$, with $\varphi_{T(B),A(B)}(r)$ being the radial wave function of the $A(B)$ sublattice at the top (bottom) layer, $q$ being the angular momentum index of the Floquet state.

Because the spatial gradient of the optical field is small, the bulk band gap of a surface element at location $\mathbf{r}$ can be estimated by diagonalization of the effective Hamiltonian at $\mathbf{r}$ with $k_{x}=k_{y}=0$, which is given as $E_{\Gamma}=2\sqrt{\Delta^{2}+V^{2}+|t_{\bot,1}|^{2}+t_{\bot,0}^{2}/2-\sqrt{\nu}/2}$ with $\nu=4|t_{\bot,1}|^{2}(4\Delta^2+t_{\bot,0}^{2})+(4\Delta V-t_{\bot,0}^{2})^{2}$. Because of the rotational symmetric, $E_{\Gamma}$ is not dependent on $\phi$. As $r\rightarrow\infty$, the optical field is near to zero, so that $E_{\Gamma}\rightarrow|2V|$. Thus, at the region with $E_{\Gamma}<|2V|$ ($E_{\Gamma}>|2V|$), the optical perturbation can be considered as effective trapping potential (effective potential barrier). At the region with effective trapping potential, the electrons can be trapped to form localized Floquet states. If the amplitude of the optical field is small, $|t_{\bot,1}|$ can be neglected. Thus, $E_{\Gamma}\approx2|V+\Delta|$. For the valley with $\tau=sign(V/\eta)$, $2|V+\Delta|<2|V|$, so that the irradiated region satisfies the trapping condition. As the amplitude of the irradiation increases, the effective trapping potential become deeper and wider, so that more localized Floquet states are induced. For the another valley with $\tau=-sign(V/\eta)$, $2|V+\Delta|>2|V|$, so that the irradiation induces effective potential barrier at the region of the light spot. However, as the amplitude of the optical field increases, the effect of $|t_{\bot,1}|$ is not negligible. Narrow regions with effective trapping potential could exist, so that a few localized Floquet states are induced. The qualitative analysis is confirmed by the numerical result in the next section.

Because the high-frequency expansion is valid under the condition $\hbar\omega>2|t_{\parallel}|$, for the BLG with $|t_{\parallel}|=2.8$ eV, the solution of Eq. \ref{equationgroup} can accurately describe the Floquet state with optical frequency being larger than $5.6$ eV. If the optical field is smaller than $5.6$ eV and larger than $2.8$ eV, the high-frequency expansion is not accurate, but the numerical result can approximately describe the Floquet states. As the optical frequency being smaller than $2.8$ eV, the high-frequency expansion is not valid. In general case, the Floquet state is described by an effective Hamiltonian, which is obtained by expanding the Floquet Hamiltonian $H_{F}=H(t)-i\hbar\frac{\partial}{\partial t}$ in the Sambe space \cite{Usaj14,sambe1,sambe2}. The effective Hamiltonian contains infinite number of Floquet replicas. In numerical simulation, the number of the Floquet replicas can be truncated as $N_{F}$, so that the effective Hamiltonian is $4N_{F}$ dimension. Therefore, a systems of $4N_{F}$ partial differential equations is required to accurately describe the Floquet state. In this paper, we focus on the cases with optical frequency being larger than $2t_{\parallel}$ by applying the high-frequency expansion.


\section{numerical result}

\begin{figure*}[tbp]
\scalebox{0.52}{\includegraphics{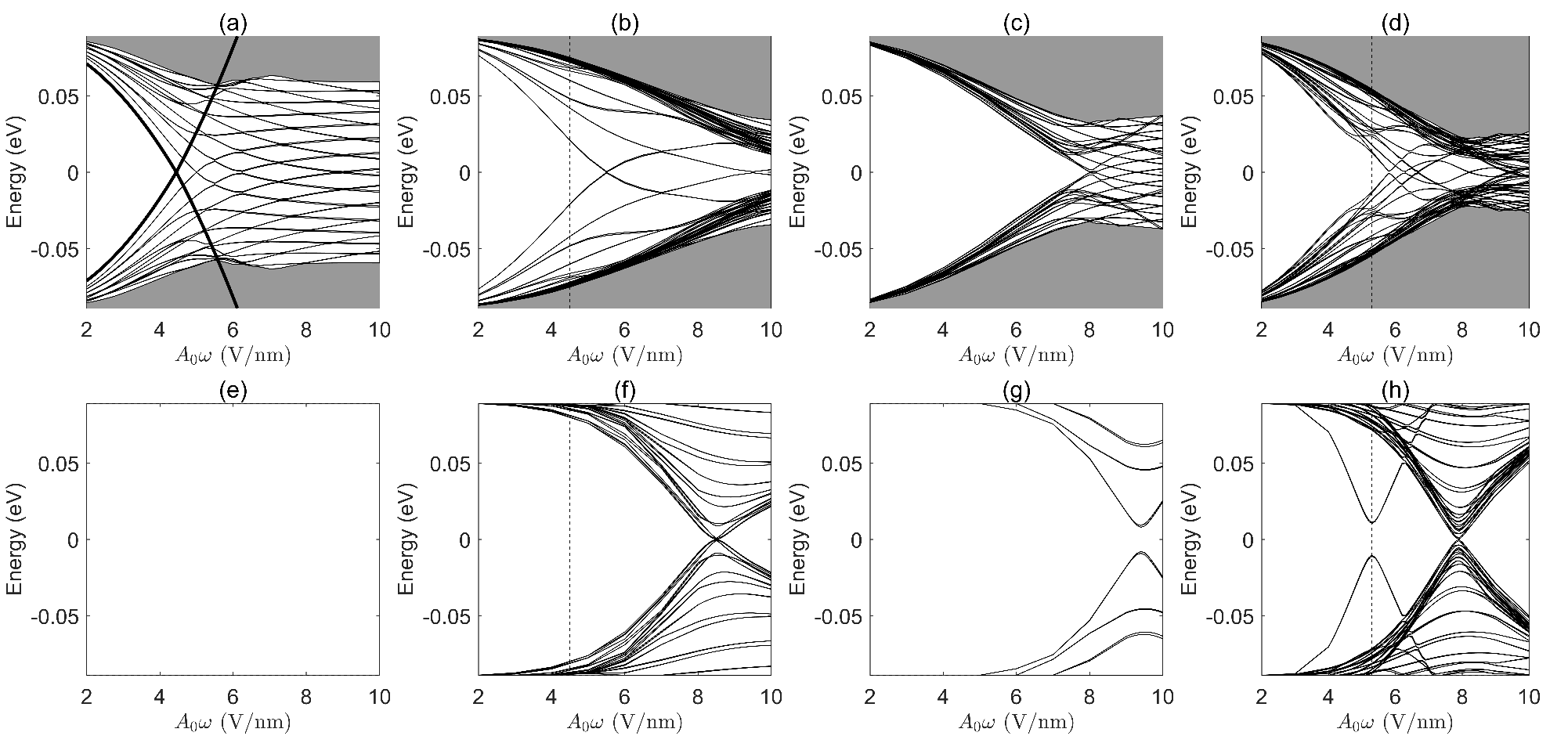}}
\caption{ The energy level of the Floquet states within the bulk band gap of the BLG, vs the amplitude of the OAM beam $A_{0}$. The radial and azimuthal indices $(p,l)$ of the OAM beam for the results in panels (a) (e), (b) (f), (c) (g), and (d) (h) are $(0,0)$, $(1,0)$, $(0,1)$ and $(1,1)$, respectively. The results of the K and K$^{\prime}$ valleys are plotted in panels(a)-(d) and (e)-(h), respectively. In panels (a)-(d), the grey area represents quasi-continuum energy spectrum. The vertical dashed line in panels (b), (d), (f), and (h) mark the system in Figs. \ref{figure_3}(a), \ref{figure_3}(b), \ref{figure_3}(c), and \ref{figure_3}(d), respectively. The thick-black line in panel (a) is the bulk band edge of the Floquet states of the BLG being irradiated by plane wave with $U=1$, $l=0$, and $p=0$ in Eq. (\ref{fieldOAM}).  }
\label{figure_1}
\end{figure*}

In our numerical simulation, the optical frequency is assumed to be $\hbar\omega=7$ eV, and the circular polarization index is assumed to be $\eta=+1$. The wavelength of the corresponding photon in vacuum is $\lambda=177$ nm. Assuming that the OAM beam is focused by metalen with numerical aperture as high as $NA=1.73$ \cite{metalen18b}, the beam waist at $z=0$ is equal to $w_{0}=\lambda/(\pi NA)=32.6$ nm. The gate voltage is assumed to be $V=0.1$ eV, so that $E_{\Gamma}=0.2$ eV. Because of the band dispersion of bulk BLG, the accurate bulk band gap in the absence of the optical field is equal to $0.178$ eV. In the presence of the optical field with varying amplitude $A_{0}$, the energy level of the Floquet states within the bulk band gap are plotted in Fig. \ref{figure_1}. For OAM beam with varying radial and azimuthal indices, the energy spectrums are different. For the four combination with $p\in[0,1]$ and $l\in[0,1]$, the corresponding energy spectrum versus $A_{0}$ are plotted in Fig. \ref{figure_1}(a-h). For the cases with $(p,l)=(0,0)$, $(0,1)$, $(1,0)$, and $(1,1)$, the values of $A_{0,\bot}\omega$ are equal to $16.9$, $9.0$, $7.4$, and $4.8$ V/nm, respectively. In the numerical calculation, 160 eigenvalues around zero energy are obtained by the iterative matrix solver. For the K valley with $\tau=+1$, the number of localized Floquet state within the band gap is much larger than 160, because large area of the irradiated region satisfies the trapping condition ($E_{\Gamma}<|2V|$). Because we are only interested in the energy level near to zero energy, the energy spectrum with large absolute value are marked by the grey area in Fig. \ref{figure_1}(a-d). For the K$^{\prime}$ valley with $\tau=-1$, the number of localized Floquet state within the band gap is smaller as shown in Fig. \ref{figure_1}(e-h), because the annular region that satisfies the trapping condition is shallow and narrow.

As $A_{0}$ reaches certain critical values, in the $\tau$ valley, one or two lowest (highest) energy levels above (below) zero energy reach zero energy. Meanwhile, in the $-\tau$ valley, the energy spectrum remains being gapped. Thus, an eight-fold or 16-fold degenerated zero energy level appears. In the $\tau$ valley, the difference between the zero energy level and the first non-zero energy level is designated as $E_{zn}^{\tau}$. In the $-\tau$ valley, the lowest energy level above zero energy is designated as $E_{g}^{-\tau}$. The degree of isolation of the zero energy level from the other energy levels is characterized by the value $E_{d}=min(E_{zn}^{\tau},E_{g}^{-\tau})$. In order to generate pure quantum states at the zero energy level for application such as quantum computation and quantum information, large $E_{d}$ is necessary. The energy spectrum of the BLG irradiated by the four types of OAM beams are discussed as following:

(i) For the system with $p=0$ and $l=0$, the energy spectrum in the K and K$^{\prime}$ valleys are plotted in Fig. \ref{figure_1}(a) and (e), respectively. As comparison, the bulk band edge of the Floquet states of the BLGs under irradiation of plane wave are plotted as thick black lines. The plane wave is equivalent to the Gaussian beam with $w_{0}\rightarrow\infty$, and then $U(\mathbf{r})\rightarrow1$ in Eq. \ref{fieldOAM}. In this case, $A_{z}(\mathbf{r},t)\rightarrow0$, so that the interlayer hopping terms are not modified. The effect of the irradiation is equivalent to a spatially uniform valley-polarized staggered sublattice potential in each graphene layer, i.e., $\Delta=-4\tau\eta(e_{0}A_{0}v_{F})^{2}/(\hbar\omega)$, so that the bulk band edge is approximately given by $E_{\pm,p}\approx\pm(V+\Delta)$. The Floquet states are non-localized, and have continuum energy spectrum above $E_{+,p}$ (below $E_{-,p}$). For the K valley, as $A_{0}$ increases from zero to a critical value $A_{0,p}$ that satisfies the condition $V-4(e_{0}A_{0,p}v_{F})^{2}/(\hbar\omega)=0$, the bulk band gap decreases and reaches zero. As $A_{0}$ further increases, the bulk band gap increases. For the K$^{\prime}$ valley, the sign of $\Delta$ is the same as $V$, so that the bulk band edge is larger than $V$. For the BLGs under irradiation of the Gaussian beam, $w_{0}$ is decreased to a finite value, then the energy spectrum is modified due to two factors. Firstly, the spatial region with $E_{\Gamma}<|2V|$ become a  circular area with finite radius, so that the continuum energy spectrum between $E_{\pm,p}$ and $\pm V$ become discrete energy levels due to finite size effect. Secondly, as $A_{0}>A_{0,p}$, the interlayer hopping terms are strongly modified, so that discrete energy levels appear between $E_{+,p}$ and $E_{-,p}$. For the systems with varying $(p,l)$, the comparison with plane wave have the same features, which can be found in Fig. \ref{figure_1}(a-h). As $A_{0}\omega$ increases from zero to $10$ V/nm, in the K valley, many energy levels emerge from the bulk band and approach zero energy; in the K$^{\prime}$ valley, no energy level emerge from the bulk band. The first zero energy level appears as $A_{0}\omega$ reaches $5$ V/nm, with $E_{zn}^{+}=0.015$ eV. The zero energy level is eight-fold degenerated. With the same $A_{0}\omega$, $E_{g}^{-}$ is much larger than $E_{zn}^{+}$, so that $E_{d}=0.015$ eV. As $A_{0}\omega$ further increases, more zero energy levels appear, whose $E_{d}$ is smaller than $0.015$ eV.

(ii) For the system with $p=1$ and $l=0$, the energy spectrum in the K and K$^{\prime}$ valleys are plotted in Fig. \ref{figure_1}(b) and (f), respectively. In the K valley, as $A_{0}\omega$ reaches $5.5$ V/nm, the first eight-fold degenerated zero energy level appears with $E_{zn}^{+}=0.027$ eV. With the same $A_{0}\omega$, $E_{g}^{-}$ is much larger than $E_{zn}^{+}$ so that $E_{d}=0.027$ eV. As $A_{0}\omega$ further increases and reaches $8.5$ V/nm, four energy bands of the K$^{\prime}$ valley cross at zero energy, so that the zero energy level is 16-fold degenerated. $E_{zn}^{-}$ of the zero energy level is $0.011$ eV. With the same $A_{0}\omega$, $E_{g}^{+}=0.009$ eV, so that $E_{d}=0.009$ eV.

(iii) For the system with $p=0$ and $l=1$, the energy spectrum in the K and K$^{\prime}$ valleys are plotted in Fig. \ref{figure_1}(c) and (g), respectively. In the K valley, as $A_{0}\omega$ exceed $8$ V/nm, a few zero energy levels with $E_{zn}^{+}<0.002$ eV appear. In the K$^{\prime}$ valley, the energy spectrum has large gap. Thus, the zero energy levels have very small $E_{d}$.

(iv) For the system with $p=1$ and $l=1$, the energy spectrum in the K and K$^{\prime}$ valleys are plotted in Fig. \ref{figure_1}(d) and (h), respectively. In the K valley, as $A_{0}\omega$ reaches $5.8$ V/nm, the first zero energy level appear with $E_{zn}^{+}=0.008$ eV, while $E_{g}^{-}$ is larger than $E_{zn}^{+}$, so that $E_{d}=0.008$ eV. The other zero energy levels have smaller $E_{d}$. In the K$^{\prime}$ valley, as $A_{0}\omega$ reaches $7.93$ V/nm, a zero energy level appear with $E_{zn}^{-}$ being very small. The energy spectrum near to zero energy level is nearly continue.

By summarizing the numerical result, all systems have zero energy levels in the K valley; the systems with $p\ne0$ ($p=0$) have (not) zero energy levels in the K$^{\prime}$ valley. For the systems with $p=0$, $L_{p}^{l}(\xi)=1$, so that $u(r)$ has not node. Thus, the absolute value of $\partial u/\partial r$ is small, and then $|t_{\bot,1}|$ is small. For the K$^{\prime}$ valley, $\Delta$ induces effective potential barrier, and $|t_{\bot,1}|$ induces effective trapping potential. The competition between the two terms generates weak effective trapping potential, so that the energy level of the localized Floquet states cannot reach zero energy. The zero energy level for the system with $p=1$, $l=0$, and $A_{0}\omega=5.5$ V/nm has the largest value of $E_{d}$. However, the zero energy level is eight-fold degenerated.

\begin{figure}[tbp]
\scalebox{0.67}{\includegraphics{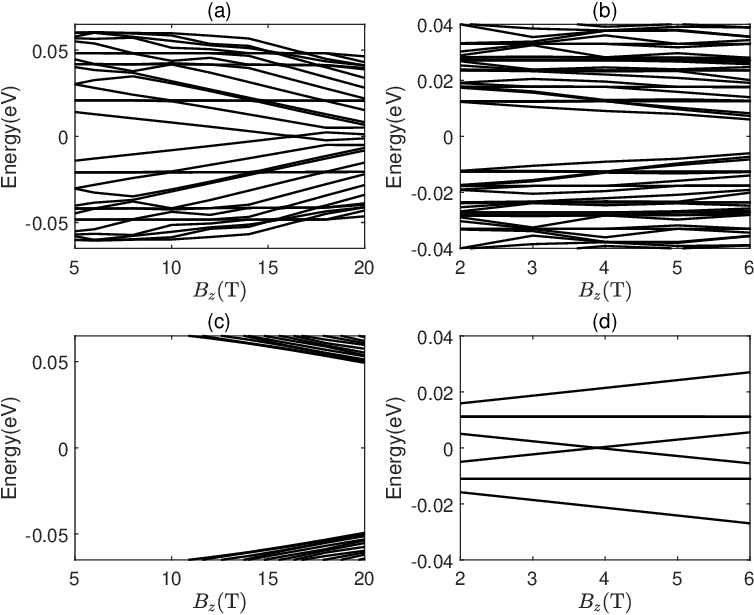}}
\caption{ The energy level versus the magnitude of the static out-of-plane magnetic field for the system in the presence of OAM beam with (a,c) $p=1$, $l=0$ and $A_{0}\omega=4.5$ V/nm, (b,d) $p=1$, $l=1$ and $A_{0}\omega=5.3$ V/nm. The results of the K and K$^{\prime}$ valleys are plotted in figures(a,b) and (c,d), respectively.    }
\label{figure_3}
\end{figure}

In the additional presence of static magnetic field, each band of four-fold degenerated energy level is split into three energy levels, with the middle energy level being two-fold degenerated, and the other two energy levels being non-degenerated. If the strength of the magnetic field is weak, the middle energy level is hardly dependent on the magnetic field. As a result, for the systems with eight-fold degenerated zero energy level, the static magnetic field split the zero energy level into three energy levels, with the zero energy level being four-fold degenerated.

\begin{figure*}[tbp]
\scalebox{0.53}{\includegraphics{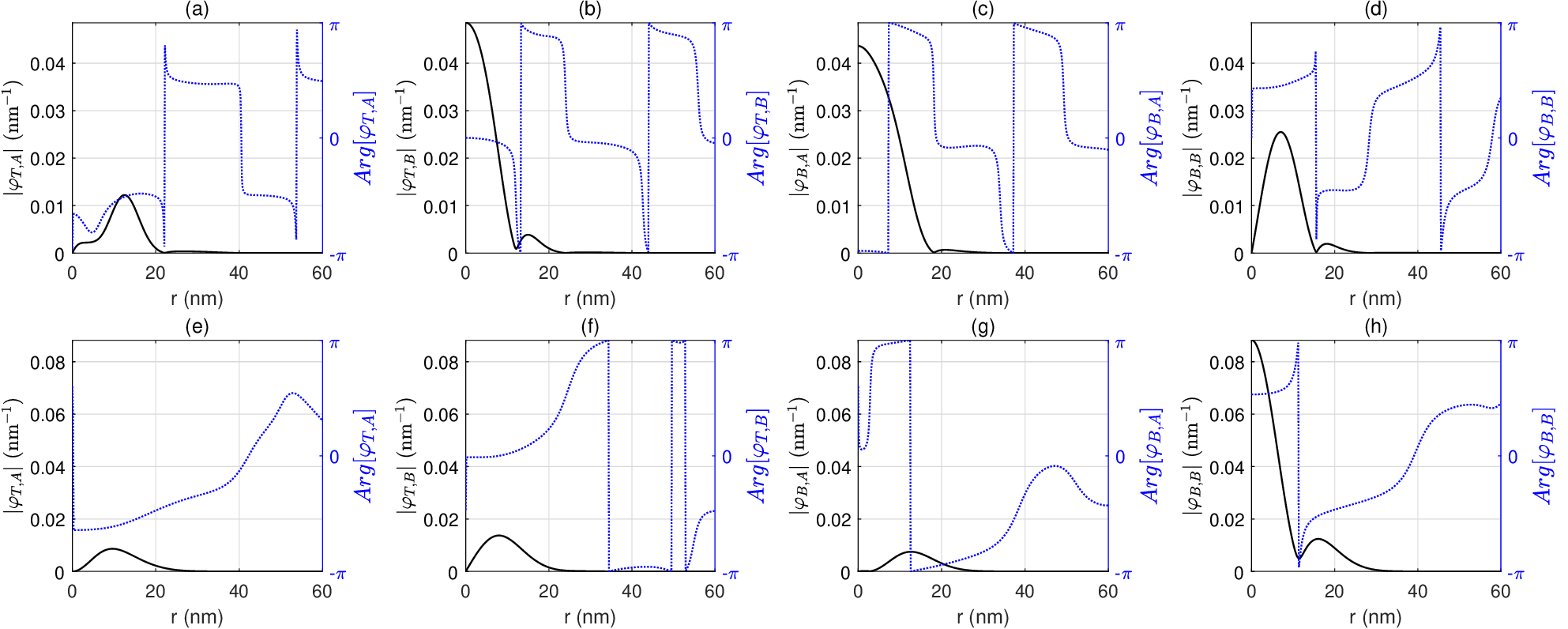}}
\caption{ The radial wave functions of the zero energy Floquet states in Fig. \ref{figure_3} (a) and (d) are plotted in the top and bottom rows, respectively. The four components of the wave function, i.e., $\varphi_{T,A}$, $\varphi_{T,B}$, $\varphi_{B,A}$, and $\varphi_{B,B}$ are plotted in the first, second, third, and fourth column, respectively. The absolute value and phase angle of the wave function are plotted as black (solid) and blue (dotted) lines, respectively.   }
\label{figure_4}
\end{figure*}

In order to induce a two-fold degenerated zero energy level, the magnetic field is applied to the irradiated BLGs, which have not zero energy level in the absence of the static magnetic field. In the $\tau$ valley, as the magnetic field split the first four-fold degenerated energy level above (below) zero energy into three energy levels, the lower (higher) energy level reaches zero at a critical magnetic field. Because each of the other four-fold degenerated energy levels are also split into three energy level at the critical magnetic field, some of the energy levels might be near to the zero energy level, so that $E_{zn}^{\tau}$ might be small. The four-fold degenerated energy levels in the $-\tau$ valley are also split, so that $E_{g}^{-\tau}$ might be also small. In order to engineer a system with two-fold degenerated zero energy level and sizable $E_{d}$, the parameters of the irradiated BLGs are selected by the following rule: in the $\tau$ valley, the first energy level above zero energy is sensitive to the static magnetic field; meanwhile, in the $-\tau$ valley, the energy gap is large, and the energy levels are not sensitive to the static magnetic field. This rule is applied to the irradiated BLGs with varying types of OAM beam, as analyzed in the following:

(i)For the system with $p=0$ and $l=0$, the energy levels in both valleys are weakly sensitive to the magnetic field, so that the critical magnetic field is larger than 25 T.

(ii) For the system with $p=1$ and $l=0$, in the K valley, the energy spectrum has a few discrete bands around zero energy as $A_{0}\omega>4$ V/nm; in the K$^{\prime}$ valley, the energy spectrum has large gap as $A_{0}\omega<6$ V/nm. Thus, as $A_{0}\omega$ being between $4$ V/nm and $6$ V/nm, applying static magnetic field could induce K-valley-polarized two-fold degenerated zero energy level with sizable $E_{d}$. The energy spectrum of a specific system with $A_{0}\omega=4.5$ V/nm versus the static magnetic field is plotted in Fig. \ref{figure_3}(a,c). As $B_{z}$ reaches a critical value $16.5$ T, in the K valley, the two-fold degenerated zero energy level appears with $E_{zn}^{+}=0.008$ eV; in the K$^{\prime}$ valley, the gap of the energy spectrum remain being large. Thus, a K-valley-polarized two-fold degenerated zero energy level with $E_{d}=0.008$ eV is induced. The radial wave functions of the four components of one of the two zero energy Floquet states are plotted in Fig. \ref{figure_4}(a-d). The angular momentum index of the zero energy Floquet states is $q=0$, so that the azimuthal wave functions of the four components are $e^{-i\phi}$, $1$, $1$, and $e^{i\phi}$. The wave functions are highly localized within a circular area with $r<w_{0}$. The total probabilities in each of the four components, i.e., $\int_{0}^{\infty}2\pi r|\varphi_{T(B),A(B)}|^{2}dr$, are equal to $0.0721$, $0.2802$, $0.4422$, and $0.2055$. Thus, the electron is mainly localized at the B sublattice of the top layer and the A sublattice of the bottom layer, which lie directly below and above each other. Near to the center of the irradiated region, the difference between the phase angle of $\varphi_{T,B}$ and $\varphi_{B,A}$ is approximately equal to $\pi$. For the another zero energy Flouqet state, the radial wave functions of the four components correspond to the four functions in Fig. \ref{figure_4}(a-d) with inverse sequence; the angular momentum index is $q=0$.

(iii) For the system with $p=0$ and $l=1$, as $A_{0}\omega>8$ V/nm, multiple energy levels in the K valley are near to zero energy , and have high sensitivity to the static magnetic field, so that $E_{zn}^{+}$ is small at the critical magnetic fields. In the K$^{\prime}$ valley, the first energy level above zero energy reaches minimum as $A_{0}\omega=9.4$ V/nm, which has large gap from the second energy level above zero energy. Thus, sizable static magnetic field could induced two-fold degenerated zero energy level with large $E_{zn}^{-}$. However, $E_{g}^{+}$ is very small, so that $E_{d}$ is very small.

(iv) For the system with $p=1$ and $l=1$, as $A_{0}\omega<5.8$ V/nm, in the K valley, the energy spectrum has sizable gap; in the K$^{\prime}$ valley, the first energy level above (below) zero energy has large gap from the second energy level above (below) zero energy. Thus, adding static magnetic field could induce zero energy level in the K$^{\prime}$ valley, while remaining sizable gap in the K valley. In order to minimized the required strength of the static magnetic field, the irradiated BLG with $A_{0}\omega=5.3$ V/nm is chosen, because the first energy level above zero energy in the K$^{\prime}$ valley reaches minimum. In the presence of the static magnetic field, the energy spectrum is plotted in Fig. \ref{figure_3}(b,d). As $B_{z}=3.88$ T, in the K$^{\prime}$ valley, the two-fold degenerated zero energy level appear with $E_{zn}^{-}=0.011$ eV; in the K valley, $E_{g}^{+}=0.01$ eV. Thus, a K$^{\prime}$-valley-polarized two-fold degenerated zero energy level with $E_{d}=0.01$ eV is induced. The radial wave functions of the four components of one of the two zero energy Floquet states are plotted in Fig. \ref{figure_4}(e-h). The angular momentum index of the zero energy Floquet states is $q=-1$, so that the azimuthal wave functions of the four components are $e^{-i2\phi}$, $e^{-i\phi}$, $e^{-i\phi}$, and $1$. The total probabilities in each of the four components are equal to $0.0455$, $0.0923$, $0.0414$, and $0.8209$. Thus, the wave function is mainly localized at the B sublattice of the bottom layer within the circular area with $r<w_{0}$. For the another zero energy Flouqet state, the radial wave functions of the four components correspond to the four functions in Fig. \ref{figure_4}(e-h) with inverse sequence; the angular momentum index is $q=1$. Thus, the two zero energy Floquet states have opposite orbital angular momentum.

For the two systems with $p=1$ and $l=0(1)$, $E_{d}$ of the valley-polarized two-fold degenerated zero energy level is larger than the thermal energy of the environment at the temperature of liquid nitrogen, i.e., $k_{B}T=0.0066$ eV with $k_{B}$ being Boltzmann constant and $T=77$ K. Thus, the zero energy Floquet states can remain being coherent at laboratory environment with regular refrigeration technology.

\section{conclusions}

In conclusion, irradiation of gated BLGs by optical Gaussian beam with OAM can induce valley-polarized localized Floquet states. The energy spectrum of the Floquet states is dependent on the parameters of the OAM beam. Additional presence of static magnetic field can further tune the energy spectrum. By engineering the parameters, valley-polarized two-fold degenerated zero energy Floquet states are induced, which has large energy interval with the other Floquet states. The zero energy Floquet states can be induced by static magnetic field that is weaker than 4 T, and can be stable in the laboratory environment with temperature being 77 K.

\begin{acknowledgments}
This project is supported by the Natural Science Foundation of Guangdong Province of China (Grant No.
2022A1515011578), the Special Projects in Key Fields of Ordinary Universities in Guangdong Province(New Generation Information Technology, Grant No. 2023ZDZX1007), the Project of Educational Commission of Guangdong Province of China (Grant No. 2021KTSCX064), and the Startup Grant at Guangdong Polytechnic Normal University (Grant No. 2021SDKYA117).
\end{acknowledgments}

\clearpage

\end{document}